# Shape and Size Resonances in Second Harmonic Generation from Plasmonic Nano-Cavities

Adi Salomon[1], Marcin Zielinski[2], Radoslaw Kolkowski[2], Joseph Zyss[2] and Yehiam Prior[1]

1 Department of chemical physics, Weizmann Institute of Science, Rehovot, Israel
2 Laboratoire de Photonique Quantique et Moleculaire, Institut d'Alembert, Ecole Normale Supérieure de Cachan, France
Yehiam.prior@weizmann.ac.il

**Abstract**

The nonlinear response of sub wavelength nano-cavities in thin metal films is investigated. We report the resonant dependence of the Second Harmonic Generation by individual triangular and square holes on shape, size and wavelength. For cavities with internal nano-corrugations, giant field enhancements are observed, making them excellent candidates for high sensitivity spectroscopy.

The linear and nonlinear optical response of metallic nano-structures is dominated by surface plasmons which are collective oscillations of their metal free electrons[1, 2]. Surface plasmons can be readily excited in metallic nanostructures with dimensions smaller than visible light wavelengths. At specific optical frequencies these collective oscillations produce large polarizabilities which reinforce the local electromagnetic (EM) field and enhance the linear and nonlinear optical response of the system[3]. Surface Enhanced Raman Scattering (SERS) is an example of nonlinear optical response that is boosted by many orders of magnitude due to the intensified EM field at or near metallic "hot spots"[4, 5]. Although SERS was observed on rough silver surfaces already in the early 1970s[6], the engineering and fabrication of structures with predefined *shapes and sizes* that give rise to enhanced local fields has been rather slow.

In addition to Raman scattering, the local field enhancement can be effectively probed by the nonlinear optical response of the medium[7, 8] Second Harmonic Generation (SHG), is the lowest order nonlinear optical effect; its generated intensity which is proportional to $(|E(\omega)^2||E(2\omega)|)^2$ (where $\omega$ is the frequency of the fundamental wavelength), is therefore very sensitive to the field enhancement[9-11]. The SHG offers two distinct advantages as a probe for local field enhancement. Firstly, within the dipole approximation, for centrosymmetric materials, the bulk contribution to the generated light vanishes[12]. A second, and no less important advantage: SHG provides information on the symmetry and tensorial properties of the local fields, not readily obtainable with spontaneous Raman scattering.

Noble metals feature a centrosymmetric face cubic centered crystal structure, and therefore SHG cannot be generated by their bulk. Yet, SHG response has already been reported from metallic surfaces and spherical nanoparticles decades ago [13-17] and attributed to excitation of surface plasmon resonances[18]. The frequency and intensity of surface plasmon resonances are determined by the geometrical properties of the nanostructures. Thus, as was previously shown, both the shape and the size of the metallic nanostructure (hole or particle) affect its polarizability and its linear optical response [19-22]. As an example, the plasmonic peak position experiences a red shift with increasing nanostructure size, and for nonspherical shapes the degeneracy is removed and new plasmonic modes are observed [23-26].

Less is known about the nonlinear behavior of metallic nanostructures and the shape and size dependence of their nonlinear response[27-35]. One possible reason for the lack of detailed data on the nonlinear response of nanoparticles is the difficulty to fabricate them with reproducible shape and size. Even the best synthesis methods available for nano particles fabrication produce a distribution of sizes and shapes and varying regularity. Nano-cavities in metal films, on the other hand, can be fabricated very precisely with resolution of 10 nm by Focused Ion Beam (FIB) milling[36]. For technical reasons, the vast majority of studies of FIB-milled nanoholes were carried out on gold samples irradiated by a standard 800 nm Ti:Sapphire fs lasers. These two facts complicated the experiments, since gold does not sustain surface plasmons at 400 nm[30, 37, 38], and behaves as a dielectric at this frequency[39]. We address both problems by using silver films and longer excitation wavelengths, thus fulfilling the condition $Re|\varepsilon(2\omega)| \gg Im(\varepsilon(2\omega))$ for excitations at the second-harmonic frequencies.

Herein, we report on the measurement of the dependence of the nonlinear response on the size and shape of individual silver nano-cavities. Following three different sets of experiments, we observe a significant enhancement of the SHG response when the fundamental wavelength matches newly evidenced dimensional resonances within the nano-cavities. Furthermore, for some nano-cavities, giant SHG signal enhancements are observed and accounted for by much finer structural corrugations at the walls of the cavity.

Consider an array of square or triangular cavities of typical side length of 100 - 300nm milled in a 200 nm thick silver film by FIB (*FEI, Helios Nano Lab 600i*). The silver film was evaporated onto a clean fused silica glass under high vacuum; its roughness and grain size were measured to be smaller than 1nm and 50nm respectively. The holes shape and size have been characterized by Scanning Electron Microscope (SEM) before and after the SHG measurements in order to identify irregular holes. Typically, for each measurement, an array of ~100 holes has been fabricated, and measured in the same experimental run to provide good statistics. The distance between the holes was about 1 μm to prevent any coupling between them[40, 41]. To have similar indices of refraction on both sides of the sample, the silver surfaces were covered by a 150 nm thick polyvinyl alcohol (PVA) layer with an average refractive index in the visible to near infrared of the order of 1.5. A SEM picture of a typical sample of an array of triangular holes is shown in Figure 1a.

The sample was illuminated by a tunable Ti:Sapphire laser (*Spectra-Physics Mai-Tai HP*, 100 fs, 80MHz, 2-10 mWatt at the entrance lens, with a fundamental incoming beam tunable between 750nm-980nm). The laser was focused through the glass using a 0.7 NA objective (×60), resulting in a spot size of about 800nm at $\lambda$ = 940 nm. The *epi*-reflected SH signal was collected by the same objective and its two perpendicular polarization components were detected by two calibrated avalanche photodiodes (APD, *PerkinElmer*)[42]. A dichroic mirror was used to block the reflected fundamental beam, and appropriate band-pass filters (*Semrock*) were used to separate and isolate the SH radiation.

A typical reflected SHG signal collected from the array is shown in figure 1b. Note that the SHG emission coming from individual holes is discrete, confirming that indeed the nano-cavities are not coupled and can be considered independently. Some of the holes, however, give rise to noticeably higher (or lower) signals. In the data presented in this paper, these holes were systematically excluded from our statistical analysis (see further below for a discussion of the holes that gave rise to giant SHG enhancement). The emitted SHG was measured for a range of hole-sizes. For each array, the average integrated SHG signal per hole was extracted by integrating the SH signal from all the holes in the array, and then further normalized by dividing the total signal by the number of considered holes.

Figures 2a and 2b depict the dependence of the SHG signal on hole size, for equilateral triangles and for squares, where the shape side length, $a$, varies from 80 – 330 nm. For both shapes, a maximum signal is observed for similar side lengths, typically of the order of one fourth to one fifth of the fundamental wavelength. We note that the SHG signal intensity does not depend on the hole area or the total circumference, but rather on more specific geometrical parameters. Furthermore, changes of about 30% in the triangular hole side length result in an increase by a factor of five in the SHG intensity. For all sizes, the SHG emission from the triangular holes is larger than that from the squares. This enhancement can be ascribed to lack of inversion symmetry in the triangular holes and to its relatively sharp corners.

While classical considerations may not be fully applicable for nano-cavities of dimension smaller than the incident wavelength, we have nevertheless run classical simulation for the field distribution inside the nano-cavities upon illumination by a plane wave. The simulations were performed by solving the full set of Maxwell's equations with the three-dimensional finite element method using the *COMSOL Multiphysics* software package. In our model, a single nano-cavity was illuminated in a direction normal to the surface of the film by a linearly polarized laser beam ($\lambda$ = 940 nm). Reflections and edge effects were properly considered. The refractive index of the medium surrounding the film and filling the aperture was set to 1.5, whereas the optical properties of silver were taken from Johnson and Christy[39]. Due to their pronounced contribution to the SHG, any observation of SH emission should be compared to the strongest local fields,[43] and this is what is plotted as solid lines in figure 2a,b for the triangular and square holes.

This strong geometrical size dependence of the SHG signal and thus of the EM field may be qualitatively understood in terms of a simple two-dimensional classical model used for macroscopic cavities of sizes larger than the optical wavelength [44]. Consider the hole as a perfect metallic resonator (cavity) in which closed orbits abiding to specular reflections on the boundaries may be evidenced. The beam accumulates a phase which must be an integer multiple of $2\pi$ upon completion of a round-trip, and including an additional $\pi$ phase shift at each reflective bounce to account for field cancellation on metallic walls. Clearly, for a given geometry, this leads to a constraint on the wavelength. The lowest (first) order modes for triangular and square cavities are illustrated in the inserts of Figure 2, and are described by the following orbits.

For equilateral triangular cavities, the simplest closed orbit is an inscribed equilateral triangle obeying:

$$\frac{2\pi}{\lambda} nL + 3\pi = 2\pi m, \text{ where } L = \frac{3a}{2} \text{ and thus } \lambda = \frac{2nL}{2m-3}.$$

For square cavities two simple types of closed orbits can be inscribed:
the first are Fabry-Pérot "bouncing ball" orbits, for which

$$\frac{2\pi}{\lambda} nL + 2\pi = 2\pi m, \text{ where } L=2a \text{ and thus } \lambda = \frac{nL}{m-1},$$

and the second are "diamond-like" orbits, for which

$$\frac{2\pi}{\lambda} nL + 4\pi = 2\pi m, \text{ where } L = 2\sqrt{2}a \text{ and thus } \lambda = \frac{nL}{m-2}.$$

Consequently, for a refractive index $n = 1.5$, and for the lowest modes $m = 2$ or $m = 1$ for the "bouncing ball" orbits, resonant conditions are expected at fundamental wavelength of $\lambda = 4.5 \cdot a$ for triangular cavities and at fundamental wavelengths $\lambda = 3 \cdot 2^{1/2} \cdot a$ and $\lambda = 3 \cdot a$ for diamond-line and bouncing ball orbits within the square cavity respectively. Thus, at these resonance frequencies, the SHG is expected to be significantly enhanced.

While classical arguments leading to this model are not a-priori expected to be valid for nano sizes, they seem to hold also for our nano-cavities. Likewise, this 2-D model is reductive in that we ignore 3-D effects that would be otherwise associated with the depth dimension of actual cylindrical nano-cavities. It is noteworthy that this model does not depend on the dielectric function of the metal; since the real part of the silver dielectric function is hardly changed at this regime and losses due to absorption are relatively small (we assume no penetration of the field into the metal). Moreover, imperfection in the fabrication of the shapes, squares in particular, have possibly caused the "bouncing ball" mode to be less stable, and therefore less visible. Clearly, a more advanced model is required to quantitatively explain these shape resonances, and such modeling is the subject of ongoing work.

To further characterize and account for the size dependence, we measured the SHG signal intensity as a function of the fundamental wavelength, over the range of 800-980 nm. Figure 3 shows the result for triangular (a) and square (b) holes, both with a nominal side length $a$ of 210 nm. The largest enhancement of the SHG emission is observed when the side length, $a$, is approximately one fourth of the fundamental wavelength, $\lambda$, with a dependence that seems to be more pronounced for the square holes, again, in accordance with the model above.

Next we explored the polarization dependence of the SHG emission from holes with different shapes and symmetry [9, 42, 45]. The collected SHG signal is split into two perpendicular polarization components which are measured by independent calibrated APDs, while the incident beam polarization is rotated by $2\pi$. Note that this method of measurement is different from the straightforward situation where the sample is rotated between fixed polarizers. If the sample is rotated, a threefold symmetric cavity shape yields a simple six-fold symmetry for the observed SHG polarization, whereas in our configuration, due to more complex projections of the plane of polarization on the measurement axes, a shape with threefold symmetry gives rise to a less intuitive four-fold symmetry of the observed SHG signal[10]. Figure 4 shows polar plots of the SHG polarization for holes with different shapes and symmetries. To better demonstrate the polarization dependence, we included several additional shapes as detailed in Figure 4. All the triangular holes (varying from 170 nm to 330 nm) yielded a four-fold pattern as expected from a three-fold symmetry cavity. To double check this point, we fabricated a shape with a more pronounced three fold symmetry (shown in inset b), and as expected, the four-fold pattern is even more pronounced. Rectangular holes give rise to a typical dipole emission pattern for various aspect ratios (2-3), while the square holes were hardly sensitive to the probing polarization of the incident beam.

The observed polar plots indicate that indeed the nano-cavity contour plays a role in determining the polarizability of the nanostructure, and non-spherical shapes lead to resonances of other plasmon modes at different directions. We note that in many previous studies, nanoholes and nanoparticles often gave a dipole pattern independently on the structure symmetry. [30, 46, 47] The characteristic octupolar polar plots, e.g. when observed for spherical particles, were assigned to retardation effects and to quadrupole excitation rather than to the shape of the metallic nanostructure. [28, 46-48] The good correlation between the structure symmetry and the polar plot is the first step towards studying interaction between the individual metallic nanostructures and also between molecules deposited on the nanostructures.

The above presented set of measurements strongly suggests shape resonances inside the hole in a way similar to the cavity modes. [35, 38] These resonance modes inside the hole should also affect the linear optical behavior of the hole. [49] For instance, extraordinary transmission (EOT) through the sub-wavelength hole array is generally attributed to resonances of surface plasmons polaritons set up by the periodicity of the hole array. [50] Calculations and experiments, however, have shown that localized mode inside the holes can also contribute to the overall transmission, [51] and that the spectral transmissions peaks were shifted by ~50 nm for the same array periodicity due to a different shape. [19, 49, 52, 53] Our results, which give direct measurements of the EM field enhanced by the nano-cavities, support these observations.

For some holes, we observed giant SHG signals which were several orders of magnitude stronger than the radiation observed from nominally identical holes. Figure 5 illustrates such a case for a set of triangular holes, all with 320 nm nominal side length. In order to account for these giant signals, we examined the relevant individual holes, and invariably found an additional finer internal structure of incomplete drilling of the metal layer, or additional roughness or finer corrugation at the walls of the holes, as is shown in the insets of Figure 5. Significant enhancements of electromagnetic fields are known to result from nanometer scale structures, and are explained in terms of resonant plasmonic excitations in these small-sized structures. The extreme sensitivity of these structures to corrugation and internal fine structure are the subject of current investigations, and may eventually lead to controlled fabrication of specific structures where strong enhancement of the electric fields may enable very sensitive detection, possibly down to the single molecule level[54-56]. The full control of the polarization properties offers unique experimental conditions not readily available in other 'surface enhanced' situations.

In conclusion we have shown experimentally that shape and size of individual sub-wavelength nano-cavities strongly affect their nonlinear behavior. An increase of one to two orders of magnitude in the emitted SHG intensity was observed when small changes in the geometrical parameters of the holes were introduced. The results strongly suggest the existence of shape resonances in metallic nanoholes, understood by a simple classical model. This, in turn, leads to relatively accurate estimates of the EM field in close proximity to the surface for a given shape. Understanding and control of shape and hole size effects on the EM field enhancement should extend the potential of these nano plasmonic cavities for SERS, data storage, sensing , optical switches and other nonlinear optical devices[57].

We acknowledge the support of the Weizmann-CNRS NaBi LEA laboratory. Part of this work is supported by a grant from the Grand Center for Sensors and Security and the James Franck Program.

**Figure 1.**

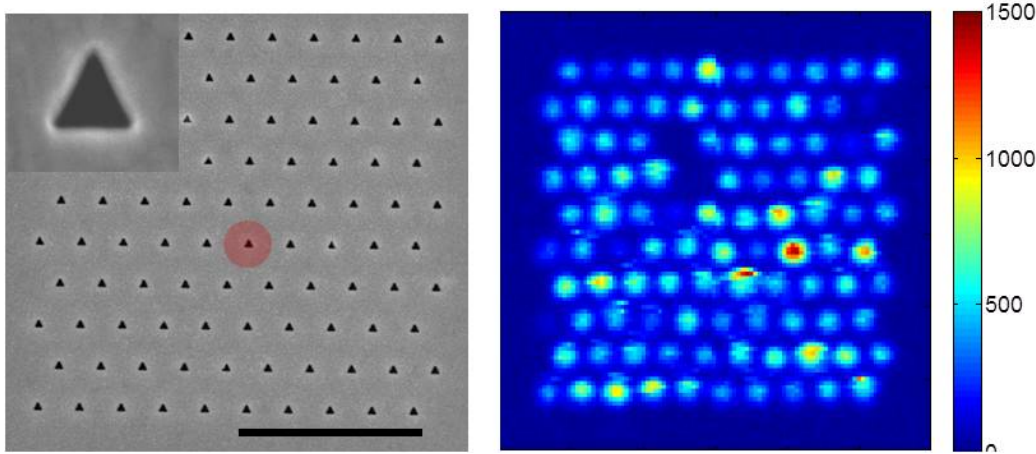

**Figure 1:** (a) SEM image of 100 equilateral holes with side length of 190 nm, separated by 1 μm. The scale bar is 5 μm. The red circle indicates the focused beam spot size in our experimental conditions. Inset: magnification of one of the holes. (b) Distribution of the SHG signal obtained from the array presented in (a) under illumination of 940 nm. The occasional low SHG signal results from blocked triangular holes as is verified by SEM.

**Figure2:**

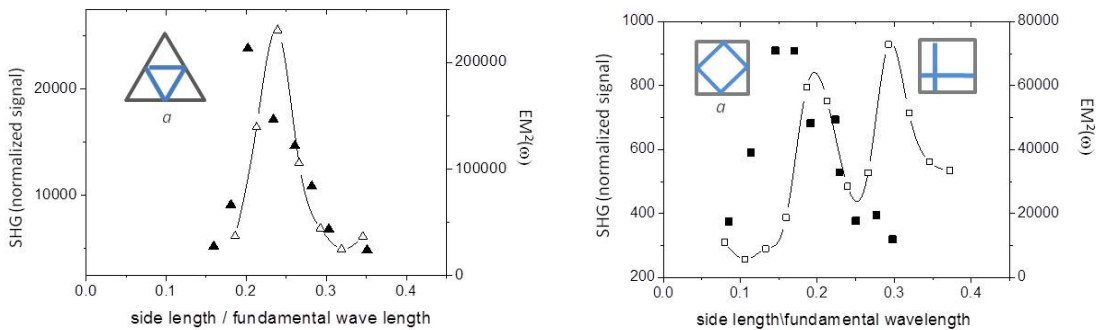

**Figure2:** (a) Normalized SHG signal (scattered triangles) for individual triangular holes (cavities) as a function of their side length divided by the fundamental wavelength. Superimposed are theoretical simulations done with *COMSOL* of the EM field stemming from the triangular holes (cavities) at the same conditions. (b) Normalized SHG signal (scattered squares) for individual square holes (cavities) as a function of their side length divided by the fundamental wavelength. Insets: examples of closed orbit modes for (a) triangular cavities and (b) square ones.

**Figure3:**

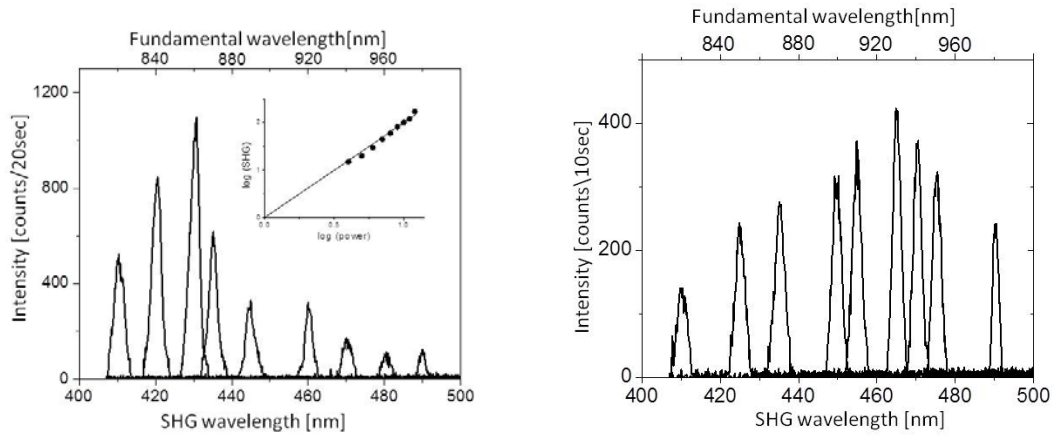

**Figure 3:** Emission spectrum from square (a) and triangular (b) nanoapertures with side length of ~210 nm. The SHG signal is divided by the bare silver surface responses and the fundamental beam power. The polarization of the incident excitation beam was set to horizontal. The fs laser power before the objective was 5 mW for triangular holes and 12 mW for square holes. Inset: the graph shows a quadratic dependence of the SHG signal on the fundamental beam power with a slope value equal 2.

**Figure 4:**

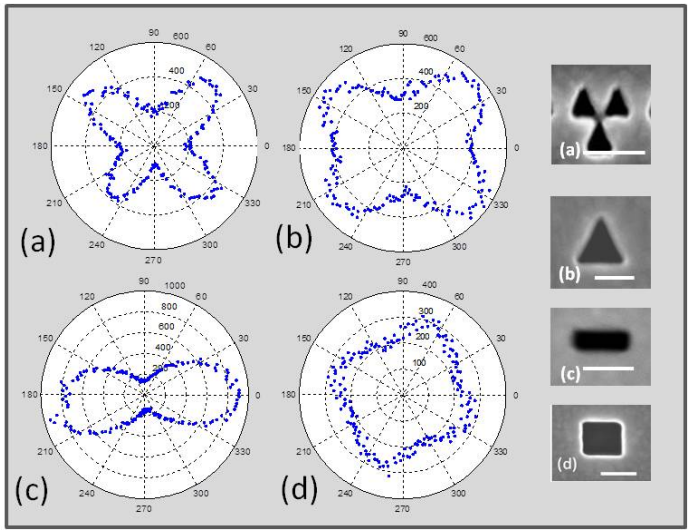

**Figure 4:** Experimental polar plots of the SH emissions for holes with different shape/symmetry. (a) A triple triangle hole – 3 fold symmetry; (b) a triangle hole; (c) a rectangular hole; (d) a square hole. The excitation wavelength was 940nm, with horizontal polarization. Each scale bar of the images at the right panel is 200 nm.

**Figure 5:**

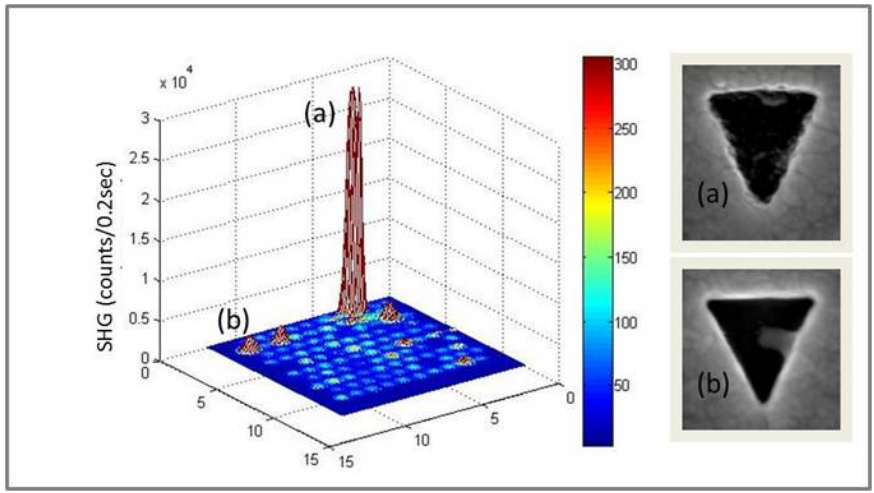

**Figure 5:** Left: enhanced SHG signals from array of isolated 330 nm side length equilateral triangles separated by 1 μm. While the average signal is about 120 counts/0.2 sec, some of the 'hot spot' triangles give rise to signals which are 200 times higher (25000 counts/0.2 sec). Right: SEM images of two hot spots as marked in the left panel. The giant SHG signal results from corrugated walls of the triangular hole.